\documentstyle[aps,epsfig,multicol,prl,psfig]{revtex}
\begin{document}
\title{Dynamics-dependent criticality in models with $q$ absorbing states}
\author{Adam  Lipowski$^{1),2)}$ and Michel
Droz$^{1)}$}
\address{$^{1)}$ Department of Physics, University of Geneva, CH 1211
Geneva 4, Switzerland\\
$^{2)}$ Department of Physics, A. Mickiewicz 
University,61-614 Pozna\'n, Poland}
\date{\today}
\maketitle
\begin{abstract}
We study a one-dimensional, nonequilibrium Potts-like model which has $q$
symmetric absorbing states.
For $q=2$, as expected, the model belongs to the parity conserving 
universality class.
For $q=3$ the critical behaviour depends on the dynamics of the model.
Under a certain dynamics it remains generically in the active phase, which is also 
the feature of some other models with three absorbing states.
However, a modified dynamics induces a parity conserving phase transition.
Relations with branching-annihilating random walk models are discussed in order
to explain such a behaviour.
\end{abstract}
\begin{multicols}{2}
\narrowtext
\section{Introduction}
Recently, dynamical and nonequilibrium properties of
many-body systems have been intensively studied.  Of particular
interest are nonequilibrium phase transitions which might appear in the
stationary state of such systems~\cite{HAYE1}.  It is believed that 
continuous phase transitions can be classified into relatively few
universality classes.  It is becoming evident, however,
that such a classification is far more complicated in non-equilibrium 
systems than in equilibrium ones.
In the equilibrium case, renormalization group and
conformal-invariance theories gave a powerful description of 
this universal phenomena. 
In non-equilibrium, the situation is far less clear due to the lack of a 
general theory. 
Accordingly, a large body of works is based on numerical simulations.

Models with absorbing states constitute a particularly rich class.  For
these models there is a substantial evidence that continuous phase
transitions can be classified into some universality classes.  In
particular, a large group of models falls into the so-called directed
percolation universality class (DP) and it was conjectured that all
models with a single absorbing state, positive one-component order parameter
and short-range dynamics should generically belong to this
universality class~\cite{DP}.  Another group consists of  models having
a double degenerate absorbing state or whose dynamics obey some
conservation law; they  belong to the parity-conserving universality
class (PC)~\cite{PC}.
 
The behaviour of models with a larger number of absorbing states was
also addressed in the literature.  For example Bassler and Browne
examined a model with three absorbing states and concluded that
depending on some parameters their model might exhibit DP or PC
criticality~\cite{BASSLER}.  However, the control parameters in their
model (adsorption rates) introduce asymmetry between absorbing states
when the critical point is approached.  Consequently,  one or two
species are effectively expelled from the system upon approaching the
critical point hence the DP or PC criticality is an expected feature of
this model.  
Similarly, certain asymmetries are responsible for the DP criticality in 
yet another model with multiple absorbing states which was studied by 
Janssen~\cite{JANSSEN}.  

To examine the role of degeneracy, one has to study models where the
symmetry between absorbing states is not broken at the level of
dynamics.  Good candidates for such a system are certain multi-species
generalizations of the contact process model~\cite{LIP96,HAYE97}.  In
the two-species case these models exhibit the expected PC criticality.
In the three-species case Hinrichsen suggested~\cite{HAYE97} that such
models will always remain in the active phase. This property, 
which should also be true for models with a larger number of absorbing states,
follows from an approximate relationship with the
$q$-species, parity conserving branching-annihilating random walk
models ($q$-BARW2, where the number at the end indicates the number of 
offsprings)~\cite{TAUBER}. 
Recently, Hinrichsen's conjecture was numerically confirmed  for models
with three and four absorbing states~\cite{CARLON}.

One can thus expect that for a given dimensionality, the 
number of (symmetric) absorbing states is the relevant parameter determining 
the critical behavior of a given model.
However, for nonequilibrium systems some other 
details of the dynamics like e.g., exclusion effects~\cite{KWON,ODOR1} or 
certain local symmetries~\cite{ODOR2}, might affect critical behaviour.
The goal of the present
paper is to provide yet another example of the dynamics-dependent criticality. 
In particular, we show that for a one-dimensional model with three absorbing 
states its critical behavior  is PC-like (instead of the
expected $q$-BARW2-like), when certain local constraints on the dynamics
of a model are introduced. 
Let us emphasize that these constraints, which can be regarded as a local
symmetry breaking, do not violate the symmetry between 
absorbing states~\cite{SYMMETRIC}.
We also suggest a mechanism which could explain such a behaviour.

Our work show that not only global properties of a given model,
like the number of absorbing states, but also certain details of the
dynamics are relevant to determine its critical behaviour.
Accordingly, classification of the critical properties of models
with absorbing states  is more complicated than originally thought.

The critical behaviour of $q$-BARW2 models was recently found to be richer than 
originally expected.
Indeed, in one dimension hard-core effects are known to change the off-critical
exponents of the model~\cite{KWON,ODOR}.
Let us notice that these effects do not change the location of the critical 
point and of the on-critical exponents~\cite{ODOR1}.
At the coarse-grained level, our model can be also regarded as a certain 
$q$-BARW2 model.
In our case, however, dynamical details have more dramatic effect: they change 
both the location of the critical point and values of all critical exponents.
\section{Model and its Monte Carlo simulations}
Before defining  our dynamical model, let us recall some basic 
properties of the usual equilibrium Potts model.
First we assign at each lattice site $i$ a $q$-state variable
$\sigma_i=0,1,...,q-1$.
Next, we define the energy of this model through the Hamiltonian:
\begin{equation}
H= -\sum_{i,j} \delta_{\sigma_i\sigma_j},
\label{e1}
\end{equation}
where summation is over pairs of $i$ and $j$ which are usually 
nearest neighbours and $\delta$ is the Kronecker delta function.
This equilibrium  model was studied using
many different analytical and numerical methods and is a rich source 
of the information about phase transitions and critical phenomena~\cite{WU}.  

To simulate numerically the equilibrium Potts model defined using the 
Hamiltonian~(\ref{e1}) one introduces a stochastic Markov process with 
transition rates chosen in such a way that the asymptotic probability
distribution is the Boltzmann distribution.
One possibility of choosing such rates is the so-called Metropolis algorithm.
In this method~\cite{BINDER} one 
looks at the energy difference $\Delta E$ between the final and initial 
configuration and accept the move with probability 
min$\{1,{\rm e}^{-\Delta E/T}\}$, where $T$ is temperature measured 
in units of the interaction constant of the Hamiltonian (\ref{e1}), which was
set to unity.
To obtain a final configuration one selects randomly a site and its state
(one out of $q$ in our case).
In the above described algorithm for $T>0$ there is always a positive
probability of leaving any given configuration (even when the final configuration
has a higher energy).
Accordingly, such a model does not have absorbing states for $T>0$.
\subsection{A-model}
To transform the standard Metropolis dynamics into the dynamics with 
absorbing-states we make the following modification:\\
\noindent
{\it Restriction A}:\\
When all neighbours of a given site are in the same state as this site,
then this site cannot change its state (at least until one of its
neighbours is changed).

In the following, this nonequilibrium model will be referred to as 
A-model. 
Obviously, any of $q$ ground states  of model (\ref{e1}) is an 
absorbing state of A-model and the dynamics does not favour any of 
the absorbing states.
Since the dynamics of our models is obtained from a modification of the Metropolis
algorithm of an equilibrium system, transition probabilities are parametrized
by temperature-like quantity $T$.
Strictly speaking, for our model the ordinary (i.e., equilibrium) 
temperature cannot be defined.
Nevertheless, we will refer to this quantity as temperature.

To study the properties of this A-model we performed Monte Carlo
simulations.  A natural characteristic of models with absorbing states
is the steady-state density of active sites $\rho$.
A given site $i$ is active when at least one of its 
neighbours is in a state different than $i$.
Otherwise the site $i$ is called nonactive.
In addition to the steady-state density we also looked at its time
dependence $\rho(t)$.  
In the active phase $\rho(t)$ converges to the
positive value while at criticality, $\rho(t)$ has a power-law
decay $\rho\sim t^{-\delta}$.
In the absorbing phase the density $\rho$ decays either faster than a 
power of $t$ or as 
power of $t$ but with a different exponent than the critical exponent $\delta$.
The unit of time is defined as a single (on average) update of each active 
site.

In addition, we used the so-called dynamic Monte Carlo method where one
sets the system in the absorbing state, locally initiate activity,
and then monitor some stochastic properties of surviving
runs~\cite{GRASSTORRE}.  The most frequently used
characteristics are the survival probability $P(t)$ that the activity
survives at least until time $t$ and the number of active sites
$N(t)$ (averaged over all runs).  At criticality these quantities are 
expected to have power-law decay: $P(t)\sim t^{-\delta'}$ and 
$N(t)\sim t^{\eta}$.  (For some
models $\delta=\delta'$, but exceptions are also known~\cite{HAYE1}).

Our simulations were made for various system sizes and we ensured that
the system was large enough so that our results are size
independent.

The simplest case is to consider  a one-dimensional chain.  However,
for $q=2$ and for any temperature $T$, this model is trivially
equivalent to the $T=0$ temperature Ising model with Metropolis
dynamics.  Indeed, in this case the allowed moves are only those which
do not increase energy and they are always accepted.  The same rule
governs the dynamics of the $T=0$ Ising chain.

\begin{figure}
\centerline{\epsfxsize=3cm 
\epsfbox{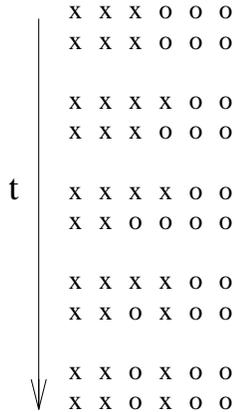}}
\vspace{5mm}
\caption{
At positive temperature a single domain can branch into additional domains.
Successive (in time) configurations differ only by a single-site flips.
States 0 and 1 are represented by "x" and "o", respectively.
In the single-chain geometry and $q=2$ such a branching is forbidden and domain 
walls can only diffuse or annihilate.
In terms of BARW models, where each domain wall represents a particle, the above 
sequence is equivalent to branching.
}
\label{figure}
\end{figure}

To overcome this geometrical 'pathology' we consider our A-model on a
ladder-like lattice where two chains are connected by inter-chain bonds
such that each site has three neighbours.
Fig.~\ref{figure} illustrates the possibility of spreading of activity in the
ladder geometry for $q=2$.
The results of  the simulations of such case are in agreement with the
up-to-date knowledge concerning one-dimensional systems with
multi-absorbing states. Thus we only briefly describe our results.  
For $q=2$ the model has two symmetric absorbing states (all $\sigma_i=0$ or 1).
Qualitatively, in this case the model resembles other models with two absorbing 
states~\cite{LIP96,HAYE97}, which are known to belong to PC universality class and
our simulations confirm such a behaviour.
\begin{figure}
\centerline{\epsfxsize=9cm 
\epsfbox{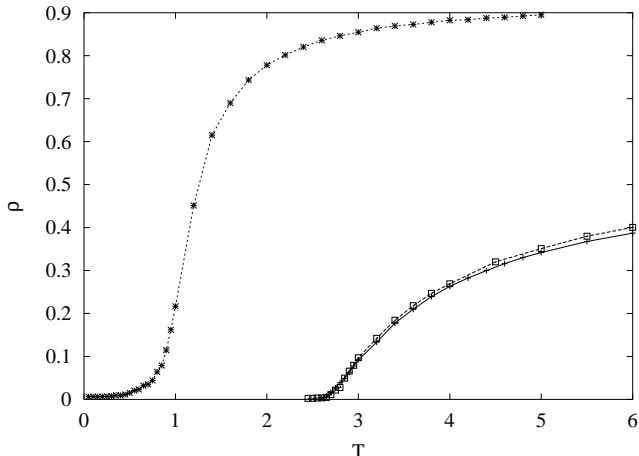}}
\caption{
The steady-state density $\rho$ as a function of $T$ for A-model
model and $q=2$ (+), 3 ($\star$) and for B-model with $q=3$
($\Box$).  Simulations were made for the system size $L=30000$ and
simulation time $t=10^5 MCS$.  The low-temperature tail for $q=3$ does not
diminish for longer simulations or larger system size.  }
\label{f1}
\end{figure}
First, the steady-state measurement of the density of active sites 
suggests a
continuous phase transition between active and absorbing phases around
$T=2.7$  (see Fig.~\ref{f1}).  Measuring the time dependence of
$\rho(t)$ we observed that at $T=2.7$ $\rho (t) \sim t^{-\delta}$ where
$\delta$ is very close to the PC value 0.286.  Moreover, in the
low-temperature ($T<2.7$) phase $\rho(t)\sim t^{-0.5}$, which is also a
typical feature of PC models.  Additional confirmation of PC
criticality in this case is obtained using the dynamical Monte Carlo
method which yields $\delta'=0.29(3)$ and $\eta=-0.02(3)$.

A different behaviour appears in the $q=3$ case.  
Although in Fig.~\ref{f1} one can see a sudden change of the order parameter
around $T=0.8$, there is no phase transition in this case.
Examining the behaviour of $\rho(t)$, we checked that even at low temperature
($T=0.5$ and 0.6) the system remains in the active phase.
In addition simulations
suggest that for $q=4$ and 5 A-model behaves similarly to the $q=3$ case.

Actually, we expect that for $q \geq 3$ the model has a critical point but 
only at $T=0$.
This point corresponds to the case of zero branching rate in a $q$-BARW2 model,
which is known to be characterized by e.g., $\beta=1$ (with the order
parameter expressed in terms of the branching rate)~\cite{CARLON}.

As we already mentioned, the absence of the transition for $q\geq 3$ and 
$T>0$ is an
expected feature.
However, as we will show below,  additional restriction in the dynamical 
rules of our model induces a transition even for $q\geq 3$. 
\subsection{B-model}
This restriction can be formulated as follows:\\
\noindent
{\it Restriction B}:\\
A flip into a state different than any of its neighbours is
prohibited.  In other words, spontaneous creation of for example 
domains of type A between domains of type B and C is forbidden.
Here, A, B, and C denote three (out of $q$) different states.

\begin{figure}
\centerline{\epsfxsize=9cm \epsfbox{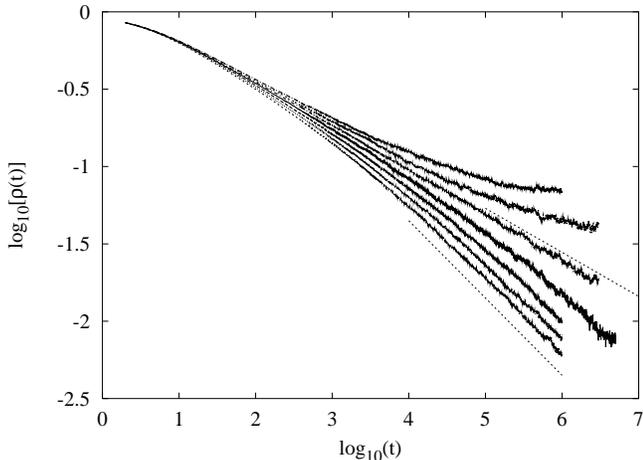}}
\caption{
The time dependence of the density $\rho(t)$ for 
B-model with $q=3$  and (from top)
$T=2.9$, 2.8, 2.7({\rm critical}), 2.6, 2.5, 2.4, and 2.3 ($L=20000$).
Each line is an average of 100 independent runs which starts from 
random initial configurations.
Straight dotted lines have slopes corresponding to $\delta=0.286$ and 0.5.
}
\label{f2}
\end{figure}

\begin{figure}
\centerline{\epsfxsize=9cm \epsfbox{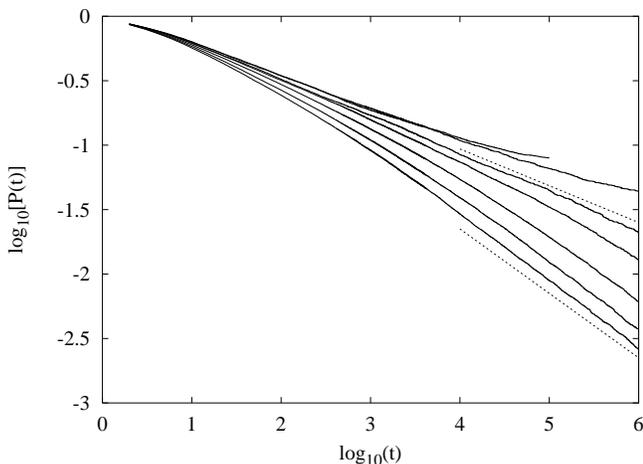}}
\caption{
The time dependence of the survival probability $P(t)$ for 
B-model with $q=3$  and (from top)
$T=2.9,\ 2.8,\ 2.7 ({\rm critical}),\ 2.6,\ 2.4, \ 2.2$, and 2.0 ($L=50000$).
Each line is an average of about $10^6$ independent runs.
Straight dotted lines have slopes corresponding to $\delta=0.286$ and 0.5 respectively.
}
\label{f3}
\end{figure}

\begin{figure}
\centerline{\epsfxsize=9cm \epsfbox{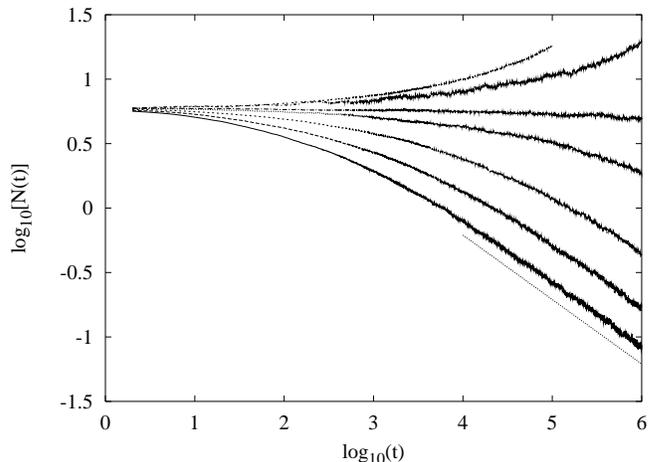}}
\caption{
The time dependence of the number of active sites $N(t)$ for 
B-model with $q=3$  and (from top)
$T=2.9,\ 2.8,\ 2.7 ({\rm critical}),\ 2.6,\ 2.4, \ 2.2$, and 2.0 ($L=50000$).
Each line is an average of about $10^6$ independent runs.
The straight dotted line has a slope corresponding to $\delta=0.5$.
}
\label{f4}
\end{figure}

Let us notice that restriction B satisfies at the same time restriction
A (but of course not {\it vice versa}).
In the following we will refer to the model satisfying restriction B as
B-model.
Let us also notice that restriction B does not break the symmetry
and B-model similarly to A-model has $q$ symmetric absorbing 
states.
Results of the simulations of B-model for $q=3$  are shown in
Fig.~\ref{f1}-Fig.~\ref{f4} (for $q=2$ both
dynamics A and B are equivalent).  First, let us notice that the density
$\rho$ (Fig.~\ref{f1}) is only slightly larger for B-model 
than for A-model with $q=2$ (the difference is, however, larger
than error bars).  In addition, this difference diminishes upon
approaching the transition point and within our numerical accuracy both
transitions seem to take place at the same temperature.  Later on we
provide some arguments which could explain this apparent coincidence.

The behaviour of $\rho(t)$ (Fig.~\ref{f2}) is typical for PC
universality class.
In the low temperature phase ($T<2.7$) we observe the power law decay 
$\rho(t)\sim t^{-1/2}$ while at criticality ($T=2.7$) 
$\rho(t)\sim t^{-\delta}$ and we estimate $\delta=0.29(2)$.
The dynamical Monte Carlo method confirms the PC criticality of this 
model~\cite{DYNAM}.
Indeed, $P(t)$ scales with the exponent $\delta'=0.29(2)$ 
(Fig~\ref{f3}) and at criticality $N(t)$ remains virtually constant 
(Fig.~\ref{f4}), which is in agreement with the PC value $\eta=0.0$.
In our opinion, the above results clearly indicate the PC criticality
of B-model.
This behavior can be qualitatively explained by considering the 
relation between our models and certain multispecies parity conserving $q$-BARW2
models.
The  relation with such models is based on the observation that
each domain wall can be at least approximately identified as a branching 
and annihilating random walker.
Although precise mappings between absorbing-states and BARW models are
usually complicated, one can argue~\cite{HAYE97} that most relevant processes 
are only long-lived ones which considerably simplifies the resulting BARW 
model.
In particular Hinrichsen argued that dynamics of models with two absorbing 
states should be approximately described by the following BARW2
model~\cite{HAYE97}:
\begin{equation}
2X \rightarrow 0,\ \  X \rightarrow 3X
\label{e2}
\end{equation}
This model is known to belong to the PC universality class~\cite{NORA}.
Since there are several types of domain walls, mappings for models with more 
than two absorbing states require several types of random walkers.
Hooyberghs et al. argued that typically in the corresponding $q$-BARW2 models the
following reactions should be included~\cite{CARLON}:
\begin{equation}
X \rightarrow Y+Z,
\label{e3}
\end{equation}
\begin{equation}
Y+Z \rightarrow X.
\label{e4}
\end{equation}
The process (\ref{e3}) represents the formation of a domain of e.g., type C between
domains of type A and B.
The process (\ref{e4}) is its reverse and corresponds to the disappearance of 
the domain C.
The $q$-species BARW2 model (\ref{e2})-(\ref{e4}) differs from the one 
studied by Cardy and T\"auber~\cite{TAUBER}.
However, the differences do not affect the critical behaviour which
is the same for both models.
We expect that our A-model for $q>2$ at the coarse-grained level is also 
described by $q$-BARW model with reactions (\ref{e2})-(\ref{e4}).

Let us now notice that in our B-model, the formation of the intruding 
domain C between domains A and B is forbidden, hence processes of the type (\ref{e3}) are 
forbidden.
The $q$-species BARW2 model which corresponds to B-model
is thus described by reactions (\ref{e2}) and (\ref{e4}).
The main point of our argument to explain the PC criticality of the B-model is 
the following:  The suppressing of the
processes (\ref{e3}) implies 
separation of time scales of
parity conserving processes (\ref{e2}), which happen at the much
shorter time scale than parity-nonconserving processes (\ref{e4}).  
As a result the $q$-BARW2 model spatially decomposes into single-species
'clouds'.
The dynamics within each 'cloud' corresponds exactly to the dynamics of
the $q=2$ B-model (which is exactly the same as as for A-model), which
at the coarse-grained level is given only by parity-conserving
processes (\ref{e2}).  Non-conserving processes (\ref{e4}) operate only when
different 'clouds' collide~\cite{COMM}.  Upon approaching the transition point
domains coarsen and the distances between 'clouds' increase.  As a result,
non-conserving processes happen only on a very-long time scale.  
(Snapshots of Monte Carlo simulations qualitatively confirm formation of such
single-species 'clouds').
On the other hand, the order parameter of the system, i.e., the number of
particles is mainly determined by the parity-conserving processes
(these processes determine the concentration of particles within
clouds).  As a result, the most relevant dynamics of the model is dominated 
by the parity-conserving processes which implies PC criticality.
As we already mentioned, these processes
correspond to the coarse-grained dynamics of the $q=2$ model which thus
explains why transition temperatures for $q=2$ and 3 are the same.
It would be interesting to confirm these qualitative considerations with more sound
theoretical arguments.
\section{Conclusions}
In conclusion, we have shown that the critical behavior of a model with 
$q$ absorbing states is not only characterized by the number of 
absorbing states but that details of the dynamics are also important.
At the coarse-grained level our model is equivalent to a certain $q$-BARW2 model.
It would be interesting to check whether suppresion of processes (\ref{e3}) in $q$-BARW2 model 
leads to a similar change of the critical behaviour.
\acknowledgements
This work was partially supported by the Swiss National Science Foundation
and the project OFES 00-0578 "COSYC OF SENS".

\end{multicols}

\begin{references}
\bibitem{DYNAM} In dynamical simulations we set all but two sites of our ladder in 
a state 0.
Two remaining sites (which are nearest neighbours) are set as 1 
and 2, respectively.
\bibitem{COMM} Reactions (\ref{e2}) and (\ref{e4}) describe our B-model only at the
coarse-grained level.
There are some other than (\ref{e4}) processes in B-model that are parity non-conserving  
but which are short lived and most likely irrelevant.
\end{references}
\end{document}